\documentclass[superscriptaddress,showpacs,preprintnumbers,amsmath,amssymb,prb]{revtex4-2}

\usepackage{graphicx}
\usepackage{bm}
\usepackage{epsfig}
\usepackage{color}
\usepackage[unicode=true,colorlinks=true,citecolor=blue,urlcolor=blue]{hyperref} 
\usepackage[utf8]{inputenc}
\usepackage[english]{babel}
\usepackage{xcolor}
\usepackage{gensymb}
\usepackage{physics}

\def\be{\begin{equation}}
\def\ee{\end{equation}}
\def\bea{\begin{eqnarray}}
\def\eea{\end{eqnarray}}

\begin{document}
\title{Extending the time of coherent optical response \\ in ensemble of singly-charged InGaAs quantum dots}

\author{A.~N.~Kosarev}
 	\affiliation{Experimentelle Physik 2, Technische Universit\"at Dortmund, 44221 Dortmund, Germany}
 	\affiliation{Ioffe Institute, 194021 St. Petersburg, Russia}

\author{A.~V.~Trifonov}
	\affiliation{Experimentelle Physik 2, Technische Universit\"at Dortmund, 44221 Dortmund, Germany}
	\affiliation{Spin Optics Laboratory, St. Petersburg State University, 198504 St. Petersburg, Russia}
	
\author{I. A. Yugova}
	\affiliation{Spin Optics Laboratory, St. Petersburg State University, 198504 St. Petersburg, Russia}
	\affiliation{V.A. Fock Institute of Physics, St. Petersburg State University, 198504 St. Petersburg, Russia}
	
\author{I.~I.~Yanibekov}
\affiliation{V.A. Fock Institute of Physics, St. Petersburg State University, 198504 St. Petersburg, Russia}

\author{S.~V.~Poltavtsev}
 	\affiliation{Experimentelle Physik 2, Technische Universit\"at Dortmund, 44221 Dortmund, Germany}
 	\affiliation{Spin Optics Laboratory, St. Petersburg State University, 198504 St. Petersburg, Russia}

\author{{A.N. Kamenskii}}
 	\affiliation{Experimentelle Physik 2, Technische Universit\"at Dortmund, 44221 Dortmund, Germany}
 	
\author{S.E. Scholz}
\author{C. Sgroi} 	
\author{A. Ludwig}
\author{A. D. Wieck}
 	\affiliation{Angewandte Festkörperphysik, Ruhr-Universit\"at Bochum, 44780 Bochum, Germany}

\author{D.R. Yakovlev}
\author{M.~Bayer}
\author{I.~A.~Akimov}
\affiliation{Experimentelle Physik 2, Technische Universit\"at Dortmund, 44221 Dortmund, Germany}
\affiliation{Ioffe Institute, 194021 St. Petersburg, Russia}

\date{\today}

\begin{abstract}
The ability to extend the time scale of the coherent optical response from large ensembles of quantum emitters is highly appealing for applications in quantum information devices. In semiconductor nanostructures, spin degrees of freedom can be used as auxiliary, powerful tools to modify the coherent optical dynamics. Here, we apply this approach to negatively charged (In,Ga)As/GaAs self-assembled quantum dots which are considered as excellent quantum emitters with robust optical coherence and high bandwidth. We study 3-pulse spin-dependent photon echoes subject to moderate transverse magnetic fields up to 1~T. We demonstrate that the timescale of coherent optical response can be extended by at least an order of magnitude by the field. Without magnetic field, the photon echo decays with $T_{\rm 2}$ = 0.45 ns which is determined by the radiative lifetime of trions $T_{\rm 1}$ = 0.27 ns. In the presence of the transverse magnetic field, the decay of the photon echo signal is given by spin dephasing time of the ensemble of resident electrons $T_{\rm 2,e} \sim$ 4 ns.  We demonstrate that the non-zero transverse $g$-factor of the heavy holes in the trion state plays a crucial role in the temporal evolution and magnetic field dependence of the long-lived photon echo signal.  
\end{abstract}

\maketitle

\section{Introduction}

The coherent optical response, which results after resonant excitation of quantum emitters with multiple optical pulses, carries rich information about the energy structure and dynamical properties of the studied system~\cite{2DFS-Cundiff,FWM-Langbein}. Moreover, it can be used for applications in quantum memories where light-matter interaction is used to store and retrieve optical fields in the form of photon echoes (PE)~\cite{Moiseev-Memory, Lvovsky-Memory, Tittel-Memory}.  In solid-state systems based on color centers and rare earth ions, significant progress has been achieved in that respect~\cite{Gisin-2011, ROSE-2011, Faraon-2017,You-zhi-2021, Moiseev-ROSE, Tittel-2021}. Yet, the search for new systems where similar or alternative approaches can be pursued on much faster times scales is of great interest~\cite{Sussman-Molecules, Sussman-Diamond, Langer-2012, Langer-2014, Kosarev-2020}.

Excitons in semiconductor nanostructures can  be addressed resonantly by sub-ps optical pulses on very short timescales enabling access to exceptionally high bandwidths, but unavoidably leading to a short radiative lifetime, which imposes limitation on the optical storage time. One of the solutions is to use the spin degrees of freedom of resident electrons in semiconductors which makes it possible to extend the timescale of coherent optical response by several orders of magnitude~\cite{Langer-2014,Salewski-2017}. The demonstration of this concept has been achieved for localized charged excitons in CdTe/(Cd,Mg)Te quantum well structures and donor-bound excitons in bulk ZnO crystals~\cite{FTT-review-2018}. It is based on resonant excitation of the donor bound exciton $D^{0}X$ or negatively charged exciton (trion) $X^{-}$ with a sequence of three resonant optical pulses in the presence of a transverse magnetic field \cite{Langer-2014}. This allows one to transfer the optical coherence of trions into the electron spin coherence of resident electrons with a significantly longer relaxation time. 

For realistic quantum memory protocols it is necessary to apply resonant optical pulses with an area of $\pi$, i.e. to perform robust Rabi flops. This is very difficult in semiconductor quantum wells and bulk crystals due to the strong damping of Rabi oscillations by excitation-induced dephasing~\cite{Langbein-2005,Poltavtsev-2019}. Moreover, weakly localized resident carriers hop between the localization sites which leads to an additional loss of the coherence~\cite{Kosarev-2019}. Therefore, it is advantageous to use quantum dots (QDs) with strong localization potential which ensures robust coherence properties~\cite{Langbein-2001,Cundiff-2016,Poltavtsev-2016,Kasprzak-2018,Kosarev-2020}. Experiments with an ensemble of QDs are challenging due to the strong inhomogeneous broadening of optical transitions. In quantum wells the optical transitions for exciton and trion are spectrally separated and, therefore, they can be selectively addressed by proper choice of the photon energy of excitation. This selectivity is not available in a QD ensemble which imposes serious restrictions for observation and subsequent application of the photon echo retrieved from resident electrons. Therefore, the demonstration of long-lived spin-dependent echoes in QDs remained unresolved. 

In this work, we demonstrate that in spite of strong inhomogeneous broadening it is possible to perform a robust transfer between the optical and spin coherence and to observe long-lived spin-dependent photon echoes (LSPE) in an ensemble of charged self-assembled QDs in a moderate transverse magnetic field. Moreover, in self-assembled (In,Ga)As/GaAs QDs the Zeeman splitting of the hole is of the same order of magnitude as that of the electron. We demonstrate that the heavy-hole splitting has a strong impact on the formation of three-pulse LSPE. In order to understand and describe properly the dynamics of LSPE in self-assembled QDs and its dependence on magnetic field, we develop a model, that accounts for both the electron and heavy-hole Zeeman splittings.

\medskip

\section{Sample and experiment}

\begin{figure*}[hbt!]
	\center{\includegraphics[width=14cm]{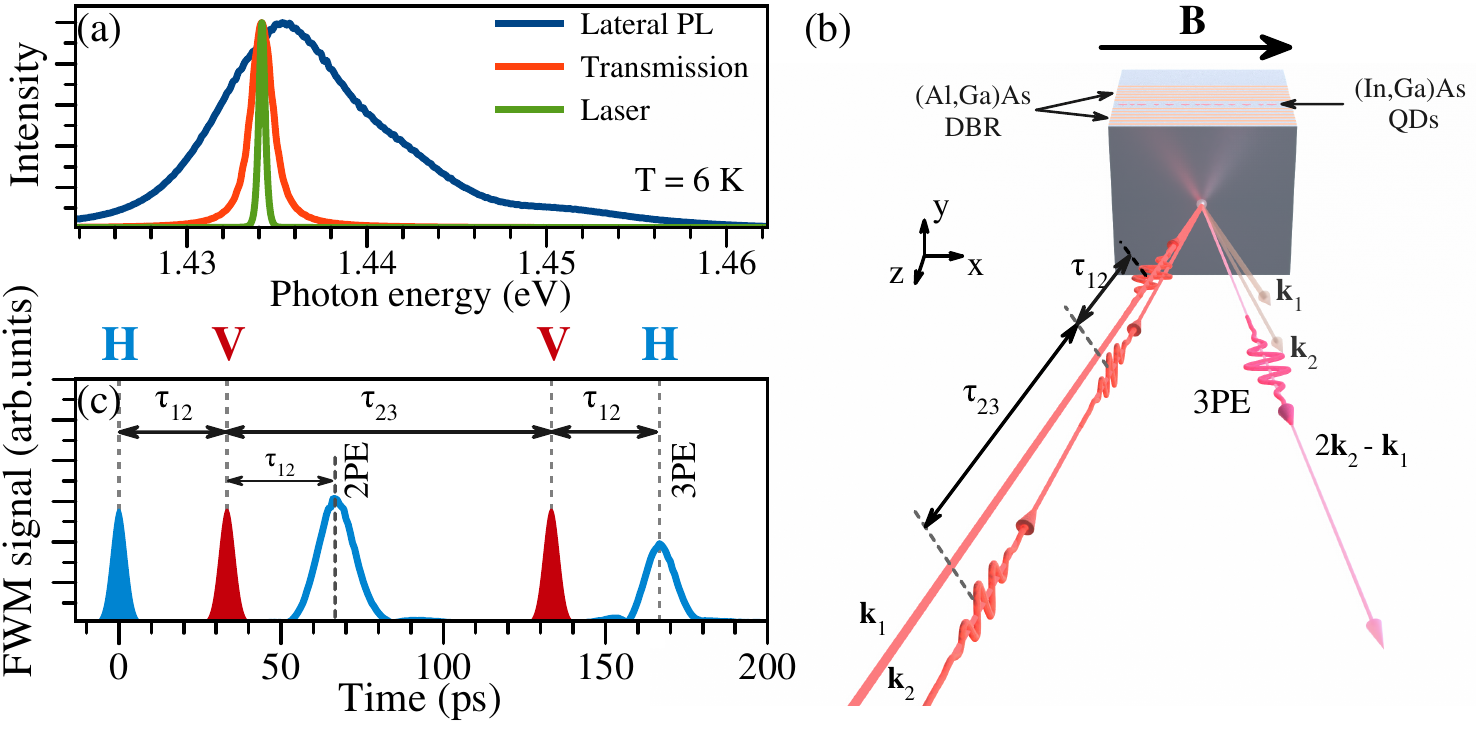}}
	\caption{{\bf Schematic representation of the experimental technique and the sample.} (a) Spectra of the sample PL, transmission and the laser. Temperature $T=6$~K. PL spectrum is shown for lateral emission from the edge of the sample in the direction parallel to its plane, e.g.  along $x$-axis. (b) Sketch of the photon echo experiment. (c) Blue line shows the transient FWM signal measured in $\mathbf k_{\rm S}  = 2 \mathbf k_{2} - \mathbf k_1$ direction for $\tau_{12}= 33.3$~ps and $\tau_{23}=100$~ps. The signal is represented by the two-pulse PE (2PE) at 67~ps and the three-pulse PE (3PE) at 167~ps. The three peaks with filled area show the temporal position of excitation laser pulses. Labels on top correspond to the polarization of excitation and detection in the HVVH configuration.} 
	\label{fig1}
\end{figure*}

The studied sample ($\#$ 14833) was grown by molecular beam epitaxy. It consists of 4 layers of n-doped (In,Ga)As QDs in GaAs matrix, embedded in the antinodes of the standing wave electric field in the microcavity. The QDs in each layer have a density of about 10$^{10}$~cm$^{-2}$. The resident electrons were supplied to the QDs by introducing $\delta$-doping with Si donors at a distance of 64.5~nm below each QD layer. After the epitaxial growth, the sample was annealed at the temperature of 900$^\circ$C to reduce the inhomogeneous broadening of the optical transitions. The QD emission is represented by the photoluminescence (PL) spectrum in Fig.~\ref{fig1}(a) which was measured from the edge of the sample in order to avoid the cavity impact (blue line). The PL maximum at the photon  energy of 1.4355~eV corresponds to the radiative recombination of excitons from the lowest confined energy state, while a weak shoulder at higher energies around 1.45 eV is apparently related to the emission from the first excited exciton states. The width of the PL line reflects the magnitude of inhomogeneous broadening for the optical transitions with the full width at the half maxima (FWHM) of 10~meV. 

The 5/2$\lambda$ microcavity is formed by 11 and 14 pairs of GaAs/AlAs layers in the top and bottom distributed Bragg reflectors, respectively, having a gradient axis in the plane of the sample along which the energy of photonic mode can be tuned. All the experiments were performed in the sample area where the photon energy of the cavity mode is in resonance with the emission peak of QDs. The corresponding transmission spectrum with a band centered at 1.434~eV and FWHM of 1.4~meV is shown by the red line in Fig.~\ref{fig1}(a). Using a microcavity with a quality factor $Q\sim$1000 facilitates the efficient generation of non-linear coherent optical signal due to the significant increase of light-matter interaction~\cite{Fras-2016,Poltavtsev-2016,Salewski-Tamm-2017}.

The sample is mounted in a liquid helium bath magneto-optical cryostat and cooled down to a temperature $T=2$~K unless stated otherwise. Laser pulses with a duration of 2.5 ps are emitted at a repetition rate of 75.75 MHz were generated by a tunable mode-locked Ti:Sapphire oscillator. The spectral width of the laser pulses with FWHM of 0.5 meV is approximately three times narrower than the photonic mode  of the cavity, i.e. the excitation pulses are not distorted by the cavity [see a the green curve in the Fig.~\ref{fig1}(a)]. The magnetic field $\mathbf{B}$ is applied parallel to the sample plane. Photon echoes are generated by a sequence of laser pulses focused into a spot of 250 $\mu$m and entering the sample under incidence close to normal with wavevectors $\mathbf k_i$ ($i$ is the pulse number, $\mathbf k_2=\mathbf k_3$), as it is shown in Fig.~\ref{fig1}(b). The pulse energy of $\mathcal{P}=5$~pJ corresponds to the pulse area of about $\pi$. The resulting transient four-wave mixing (FWM) signal is detected in reflection geometry in the direction of $\mathbf k_{\rm S}  = 2 \mathbf k_{2} - \mathbf k_1$ using heterodyne detection\cite{FWM-Langbein, FTT-review-2018}. The time-resolved electric field amplitude of the FWM signal is shown in Fig~\ref{fig1}(c) by the blue line for $\tau_{12}= 33.3$~ps and $\tau_{23}=100$~ps, where $\tau_{ij}$ is the time delay between pulses $i$ and $j$ in the sequence.  Two- and three-pulse echoes are observed at times $t=2\tau_{12}$ (2PE) and $t=2\tau_{12}+\tau_{23}$ (3PE), respectively. They are well described by Gaussian peaks with the FWHM of about 10~ps which is mainly determined by the spectral width of the excitation pulses~\cite{Kosarev-2020}. In what follows we use the magnitude of the electric field amplitude at the PE peak maximum $|P_{\rm PE}|$ to characterize the strength of the photon echo signal. 

In order to address various spin configurations, we use different linear polarization schemes in the excitation and detection paths. The direction of polarization is assigned with respect to the  magnetic field direction, i.e. H and V polarizations are  parallel and perpendicular to $\mathbf{B}$, respectively. The polarization scheme is labeled as $ABD$ or $ABCD$ for two- or three- pulse echoes. Here, the first two ($AB$) or three ($ABC$) letters indicate the linear polarizations of the optical pulses in the excitation sequence and the last letter ($D$) corresponds to the polarization direction in the detection, e.g. the data in Fig.~\ref{fig1}(c) are taken in the HVVH polarization configuration. In the case of the two-pulse PE, we used areas of pulses 1 and 2 corresponding approximately to $\pi /2$ and  $\pi$, respectively. As for the three-pulse PE experiment, we used a sequence of three $\pi /2$ pulses.
 
\section{Photon echo from trions in QDs}

\begin{figure}[hbt!]
	\center{\includegraphics[width=7cm]{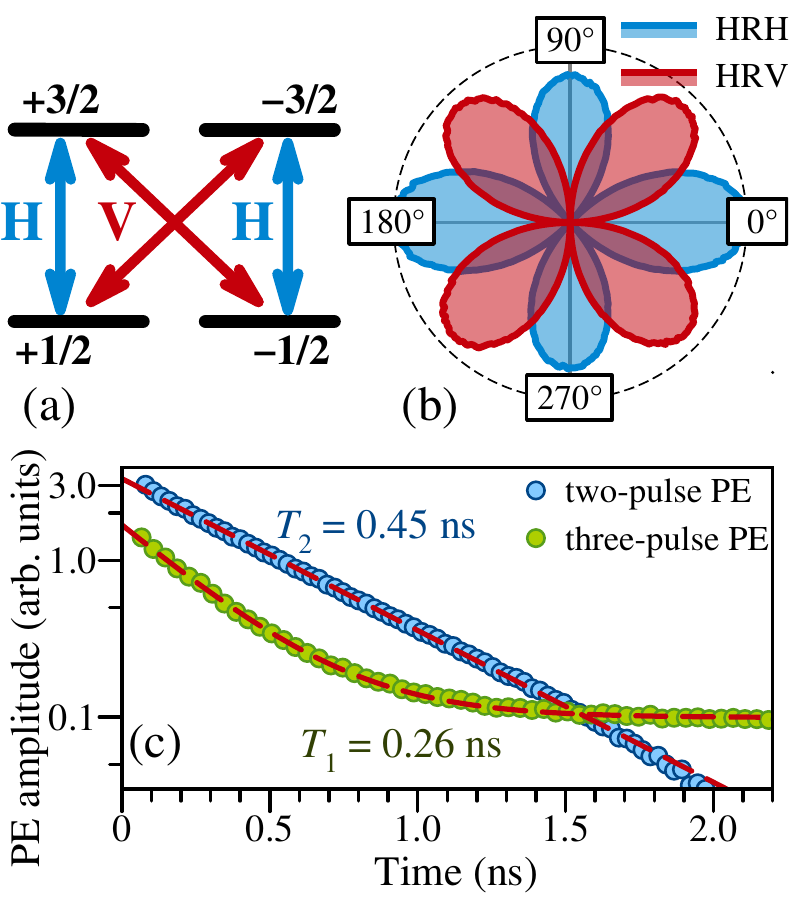}}
	\caption{ {\bf Photon echo from trions at zero magnetic field.} (a)  Energy level diagram and optical transitions for the trion $X^{-}$. (b) Polar plots of two-pulse PE amplitude in HRH and HRV polarization configurations at $t = 2 \tau_{12} = 132$ ps as function of polarization angle $\varphi_2$ of the second pulse. (c) Decay of the two- and three-pulse PE as function of $2\tau_{12}$ and $\tau_{23}$. In the three-pulse PE the delay time $\tau_{12}= 33.3$~ps. The two-pulse PE$_{12}$ decays exponentially with $T_2$ = 0.45 ns (blue circles). The three-pulse PE$_{123}$ shows exponential decay with the short time constant $T_1$ = 0.26 ns superimposed on the long-lived offset. Dashed red curves show the  corresponding exponential fits. }
	\label{fig2}
\end{figure}

In order to observe long-lived spin-dependent echoes it is necessary to address trion $X^-$ (charged exciton) complexes, which correspond to the elementary optical excitation in a charged QD. The energy spectrum in the charged QD can be well described by a four-level energy scheme with Kramers doublets in the ground and excited states at $B=0$, which are determined by the spin of the resident electron $S=1/2$ and the angular momentum of the heavy hole $J=3/2$, as shown in Fig.~\ref{fig2}(a). In contrast to the exciton in a neutral QD, this four-level scheme allows establishing optically induced long-lived spin coherence in the ground state~\cite{Salewski-2017}.

Although the photon energies for resonant excitation of trion and exciton ($X$) complexes are different in one and the same QD, it is not possible to perform selective excitation of only charged QDs by proper choice of the photon energy. This is due to the strong degree of inhomogeneous broadening for optical transitions in the QD ensemble, which is considerably larger than the energy difference between the $X$ and $X^-$ resonances. It is, however, possible to distinguish between exciton and trion contributions using polarimetric measurement of photon echo signal~\cite{Cundiff-Pola-2015,Poltavtsev-Pola-2019}. Figure~\ref{fig2}(b) shows polar plots of two -pulse PE magnitude measured at $\tau_{12}=66$~ps using HRH and HRV polarization schemes. The diagrams are obtained by rotation of the polarization direction of the second pulse (R-polarization) by angle $\varphi_2$ with respect to the H polarization.  In both polarization schemes, the signal is represented by rosettes with fourth harmonic periodicity when the angle $\varphi_2$ is scanned. Such behavior corresponds to PE response from trions where the PE is linearly polarized with the angle $\varphi_{\rm PE} = 2\varphi_2$ and the PE amplitude is independent of $\varphi_2$~\cite{Poltavtsev-Pola-2019}.  In case of the neutral exciton the polar plot is different because the PE signal is co-polarized with the second pulse ($\varphi_{\rm PE} = \varphi_2$) and it amplitude follows $|\cos\varphi_2|$. 

We note that the small increase of the PE amplitude by about 15\% in HHH as compared to HVH  remains the same under rotation of the sample around $z$-axis which excludes an anisotropy of dipole matrix elements in $xy$-plane as possible origin of asymmetry (see the blue pattern in Fig.~\ref{fig2}(b)). The difference could be provided by a weak contribution from neutral excitons. This is because in HRH configuration the PE from trions is the four-lobe pattern $\propto|\cos2\varphi_2|$ while for excitons it corresponds to a two-lobe pattern $\propto\cos^2\phi_2$. Finally, we conclude that independent of the polarization scheme the main contribution to the coherent optical response with a photon energy of 1.434~eV in the studied sample is attributed to trions. This demonstration is very important for proper interpretation of the results because long-lived spin-dependent echoes can be  observed only in charged QDs. Moreover it has large impact for applications in quantum memory protocols where high efficiency is required. 

We evaluate the optical coherence time $T_2$ and the population lifetime $T_1$ of trions in QDs from the decay of PE amplitude of the two- and three-pulse echoes, respectively. The data measured at $B=0$ in HHH polarization are shown in Fig.~\ref{fig2}(c). In the case of 2PE, the amplitude is scanned as a function of $2\tau_{12}$ (blue points), while for 3PE the dependence on $\tau_{23}$ is shown (green points). The exponential fit of two-pulse echo $|P_{\rm 2PE}| \propto \exp{(-2\tau_{12}/T_2)}$ gives $T_{\rm 2}$ = 0.45~ns which is in agreement with previous studies in (In,Ga)As/GaAs QDs \cite{Langbein-2001,Poltavtsev-2016,Kosarev-2020}. The decay of 3PE has a more complex structure. At short delay times, its magnitude decays exponentially with a time constant of $T_1=0.27$~ns which we attribute to the trion lifetime $\tau_r$. However, the signal does not decay to zero and shows a small offset with a magnitude of about 5\% of the initial amplitude at long delay times $t>1$~ns. This weak signal is governed by the dynamics of population grating in the ground state of the QDs ensemble and can be provided by many different reasons, which are out of the scope of this paper. We note that $T_{\rm 2} \approx 2 T_{\rm 1}$ indicates that the loss of optical coherence under resonant excitation of trions is governed by their radiative recombination.

\section{Long-lived spin-dependent photon echo in QDs}

Application of the transverse magnetic field ($\mathbf{B}||\mathbf{x}$) leads to Zeeman splitting of the Kramers doublets in the ground resident electron and optically excited trion states. The electron spin states with spin projections $S_x=\pm1/2$ are split by  $\hbar\omega_e = g_e\mu_B B$, while the trions states with angular momentum projections $J_x=\pm3/2$ are split by $\hbar\omega_h=g_h\mu_B B$. Here, $\omega_e$ and $\omega_h$ are the Larmor precession frequencies of electron and heavy hole spins, $g_e$ and $g_h$ are the electron and hole $g$ factor, and $\mu_B$ is the Bohr magneton. Optical transitions between all four states are
allowed using light with H or V linear polarization, as shown in Fig.~\ref{fig2}(a). The energy structure can be considered as composed of two $\Lambda$ schemes sharing common ground states. The magnetic field induces the asymmetry between these two $\Lambda$ schemes allowing one to transfer optical coherence induced by the first optical pulse into the spin coherence by application of the second optical pulse \cite{Langer-2014, Salewski-2017}. Thus, a sequence of two-linearly polarized pulses can be used to initialize a spin grating in the ground and excited states. The addressed spin components depend on the polarization of the exciting pulses. For linearly co-polarized HH sequence the spin components along the magnetic field direction are addressed (see Eq.~35 in the supplementary material) 
\begin{equation}
	\label{eq:Spins-HH}
	S_x = - J_x \propto \sin \left(  \frac{\omega_e-\omega_h }{2} \tau_{12} \right)\exp \left( - \frac{\tau_{12}}{T_2} \right)
	\cos \left( \omega_0\tau_{12} \right).
\end{equation}
In case of cross-polarized HV sequence the spin grating is produced in the plane perpendicular to the magnetic field direction (see Eqs.~36 and 37 in the supplementary material)
\begin{equation}
	\label{eq:Spins-HV}
	\begin{split}	
	S_y + iS_z=  J_y - iJ_z & \propto i \exp \left( i \frac{\omega_e-\omega_h }{2} \tau_{12} \right) \\
	&\times \exp \left( - \frac{\tau_{12}}{T_2} \right) \cos \left( \omega_0\tau_{12} \right).
	\end{split}
\end{equation}
The spectral gratings appear due to inhomogeneous broadening of the optical resonance frequencies $\omega_0$. 

The evolution of spin gratings for trions and resident electrons is governed by their population and spin dynamics. The hole spin grating lifetime is limited by the trion lifetime. The electron spin grating in the ground state is responsible for the long-lived spin-dependent echo which appears if the third pulse is applied~\cite{Langer-2014}. The decay of LSPE as a function of $\tau_{23}$ is governed by the spin dynamics of resident electrons. HHHH and HVVH polarization schemes give access to longitudinal $T_{\rm 1,e}$ and transverse $T^*_{\rm 2,e}$ spin relaxation times, respectively. 

In the studied (In,Ga)As/GaAs QDs the value of $g_h=0.18$ is of the same order of magnitude as the electronic $g$-factor $g_e=-0.52$ \cite{Trifonov-Arxiv}. Therefore, it should be taken into account in contrast to previous studies where the Zeeman splitting in the trion state was neglected. In addition, it should be noted that the PE signal depends sensitively on the orientation of crystallographic axes with respect to the magnetic field direction due to the strongly anisotropic in-plane $g$-factor of the hole in semiconductor quantum wells and QDs~\cite{Poltavtsev-PRR2020, Trifonov-Arxiv}. In our studies, the sample was oriented with the [110] crystallographic axis parallel to $\mathbf{B}$ which corresponds to the case when the H- and V- polarized optical transitions have the photon energies of $\hbar\omega_0\pm(\omega_e-\omega_h)$ and $\hbar\omega_0\pm(\omega_e+\omega_h)$, respectively.

\begin{figure*}[hbt!]
	\includegraphics[width=14cm]{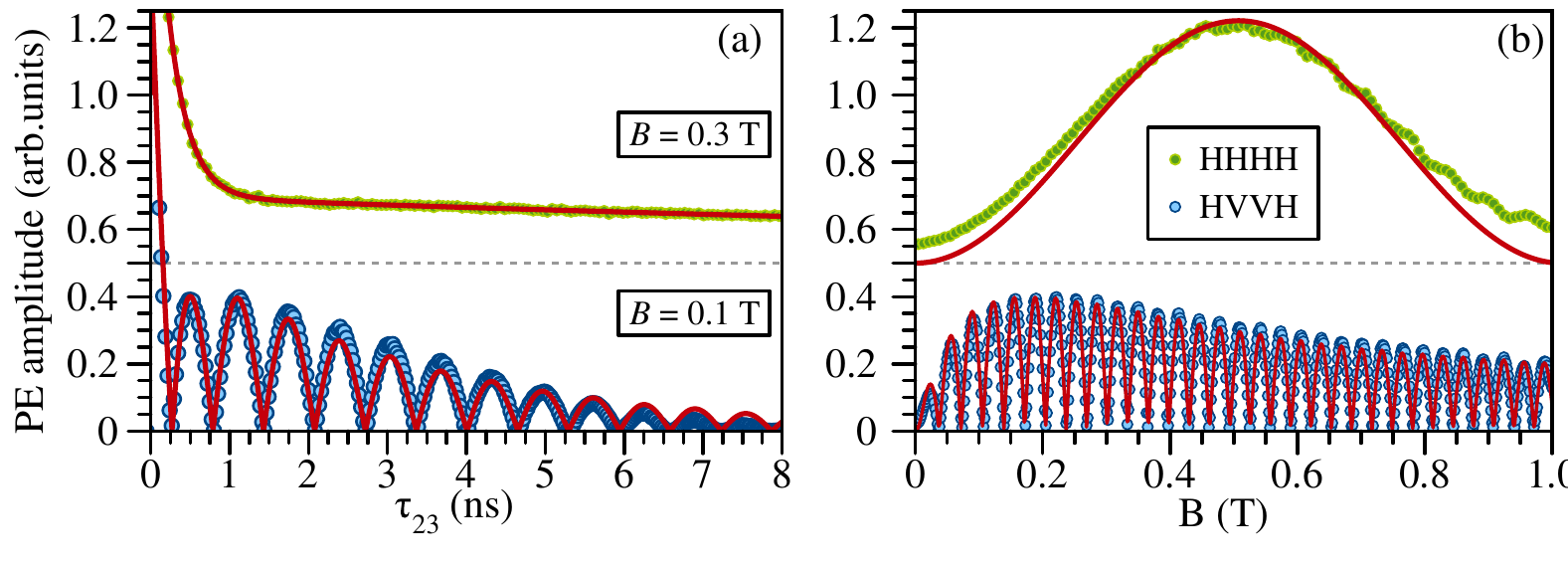}
	\center
	\caption{{\bf Long-lived spin dependent photon echo in QDs.} (a) Amplitude of three-pulse PE as a function of $\tau_{23}$ for $\tau_{12}=66$~ps. The data are taken in HHHH and HVVH polarization schemes at $B=$0.3~T and 0.1~T, respectively. (b) Magnetic field dependence of LSPE for $\tau_{12} $ = 100 ps and $\tau_{23} $ = 2.033 ns. Top and bottom curves correspond to signal measured in HHHH and HVVH polarization schemes, respectively. Red lines present the results of the theoretical modeling using Eqs.~\ref{eq:signal-HHHH} and \ref{eq:signal-HVVH} with the following parameters: $g_e= -0.516$, $g_h=0.18 $, $T_T=\tau_r=T_1=0.26$~ns, $T_{1,e}=23$~ns, $T_{2,e}^*$ is evaluated from $T_{2,e}=4.3$~ns and $\Delta g_e = 0.004$ using  Eq.~\ref{eq:T(B)} (as follows from Fig.~\ref{fig4}(a)). The signals in HHHH polarization are shifted for clarity with the dashed line corresponding to zero signal level.}
	\label{fig3}
\end{figure*}

The three-pulse PE amplitude as a function of delay time $\tau_{23}$ and magnetic field $B$ are shown in Fig.~\ref{fig3}.  In full accord with our expectations, we observe that application of a moderate magnetic field $B<1$~T drastically changes the dynamics of three-pulse PE. In HHHH polarization scheme the large offset emerges which decays on a timescale significantly longer than the repetition period of laser pulses, i.e. $T_{\rm 1,e}\gg10$~ns. The short decay, which is also present at $B=0$, with the time constant $T_1=0.26$~ns is associated to the trion lifetime. In the HVVH polarization scheme, long-lived oscillatory signal appears which is attributed to the Larmor spin precession of resident electrons and decays exponentially with $T^*_{\rm 2,e}$. At shorter delays, the signal behavior is more complex due to the superposition of spin-dependent signals from trions and resident electrons. 

Further insight can be obtained from the magnetic field dependence of LSPE signal which is measured at the long delay $\tau_{23}=2.033$~ns when the contribution from trions in three-pulse PE is negligible, see Fig.~\ref{fig3}(b). The delay time $\tau_{12}$ is set to 100~ps which is shorter than the optical coherence $T_2$.  At zero magnetic field, the PE is absent in the HVVH polarization scheme and shows  only very weak amplitude in HHHH configuration. An increase of magnetic field leads to the appearance of LSPE in both polarization configurations. For HHHH we observe a slow oscillation which is governed by Larmor precession of both electron and hole spins during $\tau_{12}$ when the spin grating is initialized by the sequence of two pulses. In the HVVH scheme the LSPE oscillates much faster because it is mainly determined by the Larmor precession of resident electron spins during $\tau_{23}$, which is roughly 20 times longer than $\tau_{12}$. 

In order to describe the experimental results quantitatively, we extended the theory from Ref. \cite{Langer-2014} by taking into account both electron and heavy-hole Zeeman splitting (for details see supplementary material). We analytically solve the Lindblad equation for the ($4 \times 4 $) density matrix to describe the temporal evolution between the first and second pulses for $0<t<\tau_{12}$ and after the third pulse for $t>\tau_{12}+\tau_{23}$. The spin dynamics of trions and electrons in external magnetic field for $\tau_{12}<t<\tau_{12}+\tau_{23}$ is described by the Bloch equations. The three-pulse PE amplitude in HHHH scheme is given by 
\begin{equation} \label{eq:signal-HHHH}
	\begin{split}
		P_{\rm HHHH}  \propto  \mathrm{e}^{-\frac{2 \tau_{12}}{T_2}} \Big[ 2 & \mathrm{e}^{-\frac{\tau_{23}}{\tau_r}} \cos^2{\left(\frac{\omega_e-\omega_h}{2} \tau_{12}\right) }  +  \\
		& \mathrm{e}^{-\frac{\tau_{23}}{T_T}}\sin^2{\left(\frac{\omega_e-\omega_h}{2} \tau_{12} \right)} + \\ & \mathrm{e}^{-\frac{\tau_{23}}{T_{1e}}}\sin^2{\left(\frac{\omega_e-\omega_h}{2} \tau_{12} \right )} \Big]
	\end{split}
\end{equation}
Here $T_T^{-1}= \tau_r^{-1} + T_h^{-1}$ is the spin lifetime of the trion. For moderate magnetic fields $B\le 1$~T we can assume that the spin relaxation time of hole in QDs $T_{h}$ is significantly longer than $\tau_r$ and, therefore, in our case $T_T=\tau_r$~\cite{Greilich-2006}. The first and second terms on the right hand side correspond to the trion contribution, while the last term is due to the LSPE from resident electrons.

For HVVH polarization we obtain
\begin{equation} \label{eq:signal-HVVH}
	\begin{split}
		P_{\rm HVVH}  \propto \mathrm{e}^{-\frac{2 \tau_{12}}{T_2}} \big[ & \mathrm{e}^{-\frac{\tau_{23}}{T_T}} r_h \cos{(\omega_h \tau_{23}-(\omega_e-\omega_h)\tau_{12}-\phi_h)}  \\
		+  & \mathrm{e}^{-\frac{\tau_{23}}{T^*_{2,e}}} r_e  \cos{(\omega_e \tau_{23}+(\omega_e-\omega_h)\tau_{12}-\phi_e)}\big] 
	\end{split}
\end{equation}
where for simplicity we introduce the following parameters: phases $\phi_e$, $\phi_h$ and amplitudes $r_e$ and $r_h$. The subscript $e,h$ corresponds to the electron or trion contributions which are given by the first and second terms on right-hand side in Eq.~\ref{eq:signal-HVVH}, respectively. The parameters are given by Eqs.~55-57 in supplementary material. They are determined by the Larmor precession frequencies $\omega_e$ and $\omega_h$, delay time $\tau_{12}$, trion lifetime $\tau_r$. The $g$-factors of electrons and holes are known from previous studies~\cite{Kamenskii-2020, Trifonov-Arxiv}. Therefore, the only unknown parameter is the spin dephasing time of resident electrons $T_{2,e}^*$. Note that if the $g$-factors of electrons and holes are unknown they can be used as additional fitting parameters in the description below.

\begin{figure}[hbt!]
	\center{\includegraphics[width=7cm]{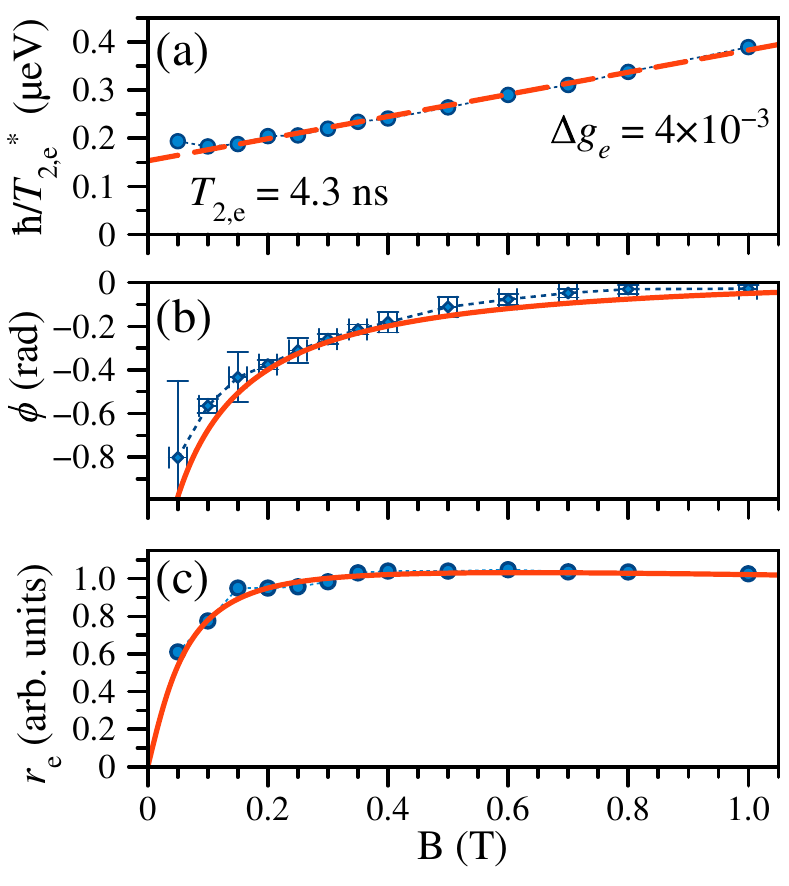}}
	\caption{{\bf Magnetic field dependence of LSPE.} Magnetic field dependence of the main parameters (decay time, phase and amplitude) of three-pulse LSPE signal evaluated from the LSPE transients $P_{\rm HVVH}(\tau_{23})$ measured at different $B$. (a) decay time $\hbar/T^*_{\mathrm 2,e}$; (b) phase $\phi$; (c) amplitude $r_e$. Blue points correspond to the data resulting from the fit using the last term on the right hand side in Eq.~\ref{eq:signal-HVVH}. Red dashed line in (a) is fit by linear function from Eq.~\ref{eq:T(B)} with $T_{2,e}=4.3$~ns and $\Delta g_e = 4\times 10^{-3}$. Red solid line in (b) and (c) - magnetic field dependences of $\phi$ and $r_e$ given by analytic expressions from supplementary material. }
	\label{fig4}
\end{figure}

In order to determine $T_{2,e}^*(B)$, we fit the transient signals in HVVH polarization for different magnetic fields as shown exemplary for the transient at $B$ = 0.1~T in Fig.~\ref{fig3}(a). For the LSPE when $\tau_{23}\gg\tau_r=2T_2$ only the second term in Eq.~\ref{eq:signal-HVVH} remains, which simplifies the fitting procedure. Three parameters of the LSPE signal, i.e. decay rate $1/T_{2,e}^*$, amplitude $r_e$, and phase $ \phi_e$, were extracted from the fit which are plotted as blue dots in Fig.~\ref{fig4} as a function of the magnetic field. It follows from Fig.~\ref{fig4}(a) that the spin dephasing rate increases linearly with the increase of $B$. Such behavior is well established in ensembles of QDs and it is related to the fluctuations of electron $g$-factor value in different QDs~\cite{Greilich-2006}. It can be described as 
\begin{equation} \label{eq:T(B)}
		\hbar/T^*_{\mathrm 2,e} = \hbar/T_{\mathrm 2,e} + \Delta g_e \mu_B B ,
\end{equation}
where $T_{\mathrm 2,e}$ is the transverse spin relaxation time and $\Delta g_e$ is the inhomogeneous broadening of the electron $g$-factor. The linear fit with this expression shown in Fig.~\ref{fig4}(a) by the red dashed line gives $T_{\mathrm 2,e} = 4.3$~ns and $ \Delta g_e = 4 \times 10^{-3}$.

The parameter $ \phi_e$ in Fig.~\ref{fig4}(b) starts from $-0.8$~rad in magnetic fields below 0.1~T and approaches zero in fields above 0.8~T. The amplitude $r_e$ in Fig.~\ref{fig4}(c) gradually rises with an increase of $B$ up to 0.4~T and remains the same in larger magnetic fields. We calculate the magnetic field dependence of amplitude and phase of LSPE using Eqs.~56 and 57 from supplementary material, respectively, using $g_e= - 0.516$, $g_h= 0.18$, $T_T=\tau_r= 0.26$~ns and $\tau_{12}= $~33.3~ps. The resulting curves are shown by red solid lines in Fig.~\ref{fig4} and are in excellent agreement with the experimental data. We note that in the limit of large magnetic fields, which corresponds to the condition of $|(\omega_e-\omega_h)|\tau_{12} \gg 1$, the amplitude of LSPE saturates ($r_e \rightarrow 1$) and the phase of the signal approaches zero ($\phi_e \rightarrow 0$) which gives the simple expression $P_{\rm HVVH}\propto \cos[\omega_e\tau_{23}+(\omega_e-\omega_h)\tau_{12}]$ for a long-lived signal at $\tau_{23} \gg \tau_r$. We emphasize that this expression takes into account the non-zero $g$-factor of the hole $g_h$ which plays an important role in the formation of the LSPE signal. 

After evaluation of $T_{2,e}^*(B)$, we can reproduce the LSPE signals as a function of $\tau_{23}$ and $B$ using Eqs.~\ref{eq:signal-HHHH} and \ref{eq:signal-HVVH} which are shown by red curves in Fig.~\ref{fig3} in both HHHH and HVVH polarization configurations. Here, the longitudinal spin relaxation rate $T_{1e}$ can be neglected, because it strongly exceeds $\tau_{23}$.  Excellent agreement is obtained at all time delays and magnetic fields. We note that the small discrepancies in HHHH polarization configuration at the magnetic fields around 0 and 1~T are attributed to the presence of a weak background signal possibly due to a population grating in the ground states as previously discussed for the case of Fig.~\ref{fig2}(c). Nevertheless, importantly the HVVH configuration which corresponds to fully coherent transformation between optical and spin coherence is free from any background.

\section{Conclusions}
In conclusion, we have demonstrated that the spin degrees of freedom can be used for substantial temporal extension of the coherent optical response in self-assembled quantum dots which has important implications for applications of this system in quantum memory devices with high bandwidth. In particular, we show that in spite of strong inhomogeneous broadening of optical transitions in the ensemble of quantum dots it is possible to store and retrieve the optical coherence in the spin ensemble of resident electrons and to extend the optical coherence time by about an order of magnitude from 0.5~ns to 4~ns. This is manifested in the emergence of long-lived spin-dependent photon echo signals under resonant excitation of trions in (In,Ga)As/GaAs quantum dots in the presence of a moderate transverse magnetic field.  We have developed a theoretical model, that quantitatively describes the behavior of three-pulse photon echo in quantum dots and takes into account the spin precession of both electrons and holes. The decay of the long-lived signal is attributed to spin dephasing of resident electrons. Therefore, the time scales can be further extended into the microsecond range by spin-echo techniques using dynamic decoupling via excitation with radio-frequency pulses. 

\section{Acknowledgements}
The authors acknowledge financial support by the Deutsche Forschungsgemeinschaft through the International Collaborative Research Centre TRR 160 (Projects A3 and A1). A.V.T. and I.A.Y. thank the Russian Foundation for Basic Research (Project No. 19-52-12046) and the Saint Petersburg State University (Grant No. 73031758). A.L. and A.D.W. gratefully acknowledge financial support from the grants DFH/UFA CDFA05-06, DFG project 383065199, and BMBF Q.Link.X 16KIS0867.




\newpage

\begin{center}
{\Large{\bf Supplementary material:\\ Extending the time of coherent optical response\\ in ensemble of singly-charged InGaAs quantum dots}}
\end{center}

\setcounter{section}{0}
\setcounter{equation}{0}
\section{introduction}

The original theoretical description of 3-pulse spin dependent photon echo (LSPE) from trions can be found in the supplementary information of Ref.~\cite{Langer-NatPhot2014}. Here we provide similar considerations but additionally take into account the spin precession of the hole. In Ref.~\cite{Langer-NatPhot2014} hole spin precession was neglected because of small hole $g$-factor in a quantum well structure. In quantum dots this is not the case and therefore it is necessary to take into account the hole precession in the photoexcited trion state. 

The description of 2- and 3-pulse photon echoes in the trion system subject to a transverse external magnetic field can be split into several steps since the duration of the optical pulses used is shorter than any characteristic relaxation times and periods of Larmor spin precession. Thus, we can neglect the relaxation processes and spin precession during the action of laser pulses, and vice versa, consider only the relaxation and evolution of the system in an external magnetic field between pulses without taking into account the interaction with light.

The following sections provide a step-by-step solution:
\begin{itemize}
\item General approach;
\item Action of optical pulses;
\item Evolution of off-diagonal matrix elements of density matrix in magnetic field
\item Evolution of diagonal matrix elements of density matrix in magnetic field;
\item Three pulse PE (3PE) signals in HHHH and HVVH polarization configurations.
\end{itemize}

\section{General approach}

	We consider an ensemble of negatively singly charged quantum dots. In this system, the ground state corresponds to a resident electron in a quantum dot, which has two possible states with spin-up $\ket{S_{\uparrow}}$ and spin-down $\ket{S_{\downarrow}}$. The optical pulse excites the quantum dots into a singlet trion state (a hole and two electrons with opposite spins in the ground state). The trion also has two possible states corresponding to the hole spin-up $\ket{J_{\uparrow}}$ and down $\ket {J_{\downarrow}}$. The eigenstates $|\pm1/2\rangle$ for electrons and $|\pm3/2\rangle$ for heavy-holes correspond to the spin projections along the main quantization axis $z$.

The Liouville equation describes the evolution of the system:
\begin{equation}
\dot{\rho} = \frac{\mathrm{i}}{\hbar} \left [H_0 + H_B + V, \rho \right] + \Gamma,
\label{LiuvEq}
\end{equation}
where $V$ is the operator of interaction with light, $H_B$ describes the action of magnetic field, $H_0$ is the Hamiltonian of unperturbed system, $\rho$ is the density matrix, and $\Gamma$ describes relaxation processes phenomenologically.

In the basis $\ket{S_{\uparrow}}$, $\ket{S_{\downarrow}}$, $\ket{J_{\uparrow}}$ and $\ket{J_{\downarrow}}$, operators $H_0$ and $H_B$ look as follows
\begin{equation}
H_0 = \begin{pmatrix}
0 &0&0&0 \\
0 &0&0&0 \\
0 &0&\hbar \omega_0 &0 \\
0 &0&0&\hbar \omega_0 \end{pmatrix}, 
\label{H0oper}
\end{equation}
\begin{equation}
H_B = \frac{\hbar}{2} \begin{pmatrix}
0 & \omega_e & 0 & 0\\
\omega_e & 0 & 0 & 0\\
0 & 0 & 0 & \omega_h\\
 0 & 0 & \omega_h & 0 \end{pmatrix},
 \label{HBoper}
 \end{equation}
where $\hbar \omega_0$ is the energy of optical transition corresponding to resonant excitation of the trion state with the lowest energy, $\omega_{e}$, $\omega_{h}$ are the electron and hole (trion) Larmor precession frequencies $\omega_e = g_e \mu_B B/\hbar$, $\omega_h = g_h \mu_B B/\hbar$.  
$\mu_B$ is the Bohr magneton.
We assume the in-plane $g$-factors of electron $g_e$ and hole $g_e$ to be isotropic.
The magnetic field $B$ is applied perpendicular to the propagation direction of the 
incident light along the $x$ axis.

The interaction with the electromagnetic wave in the dipole approximation is described
by the Hamiltonian:

\begin{equation}
 \label{eq:eq3}
\hat V(t) = -\int [\hat d_+(\bm r) E_{\sigma^+}(\bm r,t) +
\hat d_-(\bm r)E_{\sigma^-}(\bm r,t)] \mathrm d^3 r\:,
\end{equation}
where  $\hat d_\pm(\bm r)$ are the circularly polarized components of the dipole moment density operator, and $E_{\sigma ^\pm}(\bm r,t)$ are the correspondingly polarized components of the electric field of a quasi-monochromatic electromagnetic wave. The electric field of this wave is given by
\begin{equation}
\bm E(\bm r, t) = E_{\sigma^+}(\bm r,t) \bm o_+  + E_{\sigma^-}(\bm
r,t)\bm o_- + {\rm c.c.}\:, \label{eq:eq4}
\end{equation}
where $\bm o_\pm$ are the circularly polarized unit vectors that are related to the unit vectors ${\bm o}_x \parallel x$ and ${\bm o}_y \parallel y$ through $\bm o_\pm = (\bm o_x \pm \mathrm i \bm o_y)/\sqrt{2}$. Here the components $E_{\sigma ^+}$ and $E_{\sigma^-}$  contain temporal phase factors $ \mathrm e^{-\mathrm i \omega t}$.

The strength of the light interaction with the electron-trion system is characterized  by the corresponding transition matrix element of the operators $\hat{d}_{\pm}({\bm r})$ calculated with the wave functions of the valence band, $|\pm 3/2\rangle$, and conduction band, $|\pm 1/2\rangle$: \cite{ivchenko05a}
\begin{equation}
\label{dpm}
\mathsf d(\bm r) = \langle 1/2 |\hat d_- (\bm r)|3/2\rangle =
\langle - 1/2 |\hat d_+ (\bm r)|-3/2 \rangle.
\end{equation}

The Hamiltonian $\hat{V}$ in our basis is given by:
\begin{align} 
V=\frac{\hbar}{2}
	\begin{pmatrix} \label{Voper}
		0 & 0  &f_+^* \mathrm e^{\mathrm i\omega t}   &0   \\
		0 &0& 0&f_-^* \mathrm e^{\mathrm i\omega t}  \\
		f_+ \mathrm e^{-\mathrm i\omega t} &0&0 &0\\
		0&f_- \mathrm e^{-\mathrm i\omega t}  &0&0
	\end{pmatrix}.
\end{align}
 The $f_{\pm}(t)$ are proportional to the
smooth envelopes of the circularly polarized components $\sigma^+$ and
$\sigma^-$ of the excitation pulse, given by:
\[
f_{\pm}(t) = -\frac{2\mathrm e^{\mathrm i \omega t \mp \mathrm i \alpha}}{\hbar}\int \mathsf d(\bm r)
E_{\sigma_{\pm}}(\bm r,t)\mathrm d^3 r\:.
\]
Here  $\alpha$ is the angle between the $x$ axis and the polarization plane of light if it is linearly polarized. 

Relaxation processes $\Gamma$ are taken into account in the following way:
\begin{equation}
\Gamma = \begin{pmatrix}
 - \frac{\rho_{11} - \rho_{22}}{2 T_{2e}} + \frac{\rho_{33}}{\tau_r} & - \frac{\rho_{12}+\rho_{21}}{2 T_{1e}}-\frac{\rho_{12}-\rho_{21}}{2 T_{2e}} & -\frac{\rho_{13}}{T_2} & -\frac{\rho_{14}}{T_2} \\
- \frac{\rho_{12}+\rho_{21}}{2 T_{1e}}+\frac{\rho_{12}-\rho_{21}}{2 T_{2e}}  & - \frac{\rho_{22} - \rho_{11}}{2 T_{2e}} + \frac{\rho_{44}}{\tau_r} &  -\frac{\rho_{23}}{T_2} & -\frac{\rho_{24}}{T_2} \\
  -\frac{\rho_{31}}{T_2} & -\frac{\rho_{32}}{T_2} & - \frac{\rho_{33} - \rho_{44}}{2 T_{2h}} - \frac{\rho_{33}}{\tau_r} & - \frac{\rho_{34}+\rho_{43}}{2 T_{1h}}-\frac{\rho_{34}-\rho_{43}}{2 T_{2h}}- \frac{\rho_{34}}{\tau_r} \\
   -\frac{\rho_{41}}{T_2} & -\frac{\rho_{42}}{T_2} & - \frac{\rho_{34}+\rho_{43}}{2 T_{1h}}+\frac{\rho_{34}-\rho_{43}}{2 T_{2h}}- \frac{\rho_{43}}{\tau_r}& - \frac{\rho_{44} - \rho_{33}}{2 T_{2h}} - \frac{\rho_{44}}{\tau_r} \end{pmatrix}
   \label{Decayoper}
\end{equation}
Here, $T_2$ is the optical coherence decay time, $\tau_r$ is the trion recombination time, $T_{1e(1h)}$ is the electron (hole) longitidual spin relaxation time and $T_{2e(2h)} $ is the electron (hole) transverse spin relaxation time. We assumed that $T_2/2=T_1=\tau_r$ and the magnetic field is directed along the $x$~axis.  

The Liouville equation~\ref{LiuvEq} with operators~\ref{H0oper}, \ref{HBoper}, \ref{Voper} and relaxation \ref{Decayoper} is written for the density matrix $\hat{\rho}$, which in our basis is a 4x4 matrix:
\begin{equation}
\rho = \left( \begin{matrix}
\rho_{11} & \rho_{12} &\rho_{13} &\rho_{14} \\
\rho_{21} & \rho_{22} &\rho_{23} &\rho_{23} \\
\rho_{31} & \rho_{32} &\rho_{33} &\rho_{34} \\
\rho_{41} & \rho_{42} &\rho_{43} &\rho_{44} 
\end{matrix} \right).
\end{equation}

For description of the density matrix evolution in a magnetic field, it is more convenient to consider separately the evolution of (i) off-diagonal elements of the density matrix $\rho_{13}, \rho_{14}, \rho_{23}, \rho_{24} $ (+ c.c.) which are associated with optical coherences and (ii) diagonal elements $\rho_{11}, \rho_{22}, \rho_{33}, \rho_{44}$ (populations  of electron and trion spin states) in combination with off-diagonal elements $\rho_{12}, \rho_{21}, \rho_{34}, \rho_{43}$, which are associated with spin coherences of trions and electrons. This splitting has the following physical reasoning: The first category of off-diagonal elements $\rho_{13}, \rho_{31}, \rho_{24}, \rho_{42}$ are responsible for the optically induced macroscopic polarization $ P = \Trace { [\hat{d}, \hat{\rho}]} \sim \rho_{31} d_ + + \rho_{42} d_- + c.c. $, while the second category of  components are responsible for the spin state of the carriers.

\subsection{Action of short optical pulse} 

First we consider the effect of photoexcitation by a short laser pulse with frequency $\omega$ close to the trion resonant frequency $\omega_0$. For simplicity we consider optical pulses with rectangular shape, which allow us to get analytical solutions for the density matrix. 

The solution of the von Neumann equation $i\hbar \dot{\rho}=[\hat{H}_0+\hat{V},\rho]$ gives:
\bea
\rho_{11}(t_p)&=&|K_+|^2\rho_{11}(0)+|\theta_+|^2L_+^2\rho_{33}(0)+iL_+[\theta_+K_+\rho_{13}(0)-\theta_+^*K_+^*\rho_{31}(0)]\nonumber\\
\rho_{33}(t_p)&=&|K_+|^2\rho_{33}(0)+|\theta_+|^2L_+^2\rho_{11}(0)-iL_+[\theta_+K_+\rho_{13}(0)-\theta_+^*K_+^*\rho_{31}(0)]\nonumber\\
\rho_{13}(t_p)&=&e^{i\omega t_p}[K_+^2\rho_{13}(0)-i\theta_+^*L_+K_+[\rho_{33}(0)-\rho_{11}(0)]+(\theta_+^*)^2L_+^2\rho_{31}(0)]
\label{eq:eq5a}
\eea
\bea
\rho_{22}(t_p)&=&|K_-|^2\rho_{22}(0)+|\theta_-|^2L_-^2\rho_{44}(0)+iL_-[\theta_-K_-\rho_{24}(0)-\theta_-^*K_-^*\rho_{42}(0)]\nonumber\\
\rho_{44}(t_p)&=&|K_-|^2\rho_{44}(0)+|\theta_-|^2L_-^2\rho_{22}(0)-iL_-[\theta_-K_-\rho_{24}(0)-\theta_-^*K_-^*\rho_{42}(0)]\nonumber\\
\rho_{24}(t_p)&=&e^{i\omega t_p}[K_-^2\rho_{24}(0)-i\theta_-^*L_-K_-[\rho_{44}(0)-\rho_{22}(0)]+(\theta_-^*)^2L_-^2\rho_{42}(0)] 
\label{eq:eq6a}
\eea

Other off-diagonal components for elliptically polarized excitation:
\bea
\rho_{14}(t_p)&=&e^{i\omega t_p}[K_+K_-\rho_{14}(0)+\theta_+^*\theta_-^*L_+L_-\rho_{32}(0)+i\theta_-^*L_-K_+\rho_{12}(0)-i\theta_+^*L_+K_-\rho_{34}(0)]
\nonumber\\
\rho_{32}(t_p)&=&e^{-i\omega t_p}[K_+^*K_-^*\rho_{32}(0)+\theta_+\theta_-L_+L_-\rho_{14}(0)+i\theta_-L_-K_+^*\rho_{34}(0)-i\theta_+L_+K_-^*\rho_{12}(0)]
\nonumber\\
\rho_{12}(t_p)&=&K_+K_-^*\rho_{12}(0)+\theta_+^*\theta_-L_+L_-\rho_{34}(0)+i\theta_-L_-K_+\rho_{14}(0)-i\theta_+^*L_+K_-^*\rho_{32}(0)
\nonumber\\
\rho_{34}(t_p)&=&K_+^*K_-\rho_{34}(0)+\theta_+\theta_-^*L_+L_-\rho_{12}(0)+i\theta_-^*L_-K_+^*\rho_{32}(0)-i\theta_+L_+K_-\rho_{14}(0)
\label{eq:eq7a}
\eea
Here $\rho_{11}(0)$, $\rho_{33}(0)$, $\rho_{13}(0)$, $\rho_{32}(0)$, $\rho_{12}(0)$ and so on are initial conditions for the components of the density matrix. These values are equal to the corresponding matrix elements before pulse arrival. 
$\theta_{\pm}$ are pulse areas, $\theta_{\pm}=f_{\pm}t_p$,  $t_p$ is pulse duration. $$K_{\pm}=\cos(\Omega_{\pm}/2)-i \frac{\Delta}{\Omega_{\pm}} \sin(\Omega_{\pm}/2),$$ $$L_{\pm}=\frac{\sin(\Omega_{\pm}/2)}{\Omega_{\pm}}.$$ $\Delta = \omega - \omega_0$ is the
detuning between the pump frequency and the trion resonance frequency. 
$\Omega_{\pm}=\sqrt{|\theta_{\pm}|^2+(\Delta t_p)^2}$.

It is worth noting that if the excitation is linearly polarized, then $K_+=K_-$, $L_+=L_-$, and $\theta_{\pm}$ can be written (see eq.(7)) as $\theta_{\pm}=\theta_0 \mathrm e^{\mp \alpha}$.

\section{Evolution of off-diagonal matrix elements of density matrix in transverse magnetic field.}

The density matrix elements associated with optical coherence in transverse magnetic field taking into account the relaxation are given by
\bea
\rho_{13}(t) &=& \left(\rho_{13}(0)\cos(\frac{\omega_e t}{2})\cos(\frac{\omega_h t}{2})+i\rho_{14}(0)\cos(\frac{\omega_e t}{2})\sin(\frac{\omega_h t}{2})\right.\nonumber\\
&-&\left. i\rho_{23}(0)\sin(\frac{\omega_e t}{2})\cos(\frac{\omega_h t}{2})+\rho_{24}(0)\sin(\frac{\omega_e t}{2})\sin(\frac{\omega_h t}{2})\right)\mathrm e^{-t/T_2}\mathrm e^{i\omega_0 t} ,\nonumber\\
\rho_{14}(t) &=& \left(i\rho_{13}(0)\cos(\frac{\omega_e t}{2})\sin(\frac{\omega_h t}{2})+\rho_{14}(0)\cos(\frac{\omega_e t}{2})\cos(\frac{\omega_h t}{2})\right.\nonumber\\
&+&\left. \rho_{23}(0)\sin(\frac{\omega_e t}{2})\sin(\frac{\omega_h t}{2})-i\rho_{24}(0)\sin(\frac{\omega_e t}{2})\cos(\frac{\omega_h t}{2})\right)\mathrm e^{-t/T_2}\mathrm e^{i\omega_0 t} ,\nonumber\\
\rho_{23}(t) &=& \left(-i\rho_{13}(0)\sin(\frac{\omega_e t}{2})\cos(\frac{\omega_h t}{2})+\rho_{14}(0)\sin(\frac{\omega_e t}{2})\sin(\frac{\omega_h t}{2})\right.\nonumber\\
&+&\left. \rho_{23}(0)\cos(\frac{\omega_e t}{2})\cos(\frac{\omega_h t}{2})+i\rho_{24}(0)\cos(\frac{\omega_e t}{2})\sin(\frac{\omega_h t}{2})\right)\mathrm e^{-t/T_2}\mathrm e^{i\omega_0 t} ,\nonumber\\
\rho_{24}(t) &=& \left(\rho_{13}(0)\sin(\frac{\omega_e t}{2})\sin(\frac{\omega_h t}{2})-i\rho_{14}(0)\sin(\frac{\omega_e t}{2})\cos(\frac{\omega_h t}{2})\right.\nonumber\\
&+&\left. i\rho_{23}(0)\cos(\frac{\omega_e t}{2})\sin(\frac{\omega_h t}{2})+\rho_{24}(0)\cos(\frac{\omega_e t}{2})\cos(\frac{\omega_h t}{2})\right)\mathrm e^{-t/T_2}\mathrm e^{i\omega_0 t} ,\nonumber\\
\label{eq:eq11a}
\eea
where $\rho_{ij}(0)$ describes initial state of the system (initial conditions). 

The remaining non-diagonal elements of the density matrix can be obtained from $\rho_{31} = \rho_{13}^{*}$,  $\rho_{32} = \rho_{23}^{*}$,  $\rho_{41} = \rho_{14}^{*}$,  $\rho_{42} = \rho_{24}^{*}$.

\section{Evolution of spin polarization in magnetic field}

The correspondence between spin of electron and trion and density matrix elements is the following:

\begin{table}[h]
\begin{tabular}{ccc}
$S_z = (\rho_{11}-\rho_{22})/2$, & $S_y = \mathrm{i} (\rho_{12}-\rho_{21})/2$, & $S_x = (\rho_{12}+\rho_{21})/2$ \\
$J_z = (\rho_{33}-\rho_{44})/2$, & $J_y = \mathrm{i} (\rho_{34}-\rho_{43})/2$, & $J_x = (\rho_{34}+\rho_{43})/2$ \\
$n_e = (\rho_{11}+\rho_{22})$, & $n_T = (\rho_{33}+\rho_{44})$ & 
\end{tabular}
\end{table}

The time evolutions of spins are described by the following equations:
\begin{equation}
\frac{d\vec{J}}{dt} = \frac{\mu_B}{\hbar} \left [  g_h \vec{B}\times \vec{J} \right] - \frac{\vec{J}}{T_h} - \frac{\vec{J}}{\tau_r},
\end{equation}
\begin{equation}
\frac{d\vec{S}}{dt} = \frac{\mu_B}{\hbar} \left [  g_e \vec{B}\times \vec{S} \right] - S_x\frac{\vec{e_x}}{T_{1e}} -S_y\frac{\vec{e_y}}{T_{2e}} - S_z\frac{\vec{e_z}}{T_{2e}}+ J_z\frac{\vec{e_z}}{\tau_r},
\end{equation}
Here we again assumed that the magnetic field is applied along $x$-axis. We also neglected the anisotropy of hole's spin relaxation $T_h = T_{1h} = T_{2h}$ because both transverse and longitudinal hole spin relaxation are much longer then $\tau_r$.

Introducing the hole spin life time in the trion state  $1/T_T = 1/T_h+1/\tau_r$, one can obtain solution for hole time evolution:
\begin{equation}
n_T(t) = n_T(0) \mathrm{e}^{-\frac{t}{\tau_r}}
\end{equation}
\begin{equation}
J_x(t) = J_x(0) \mathrm{e}^{-\frac{t}{T_T}}
\end{equation}
\begin{equation}
J_y(t) = \mathrm{e}^{-\frac{t}{T_T}} ( J_y(0) \cos{\omega_h t} -J_z(0) \sin{\omega_h t})
\end{equation}
\begin{equation}
J_z(t) = \mathrm{e}^{-\frac{t}{T_T}} ( J_z(0) \cos{\omega_h t} +J_y(0) \sin{\omega_h t})
\end{equation}

Introducing $\gamma = 1/T_T-1/T_{2e}$ one can obtain the solution for electron spin and population dynamics:

\begin{equation}
n_e(t) = n_e {(0)} + n_T(0)\left(1- \mathrm{e}^{-\frac{t}{\tau_r}} \right)
\end{equation}
\begin{equation}
S_x(t) = S_x(0) \mathrm{e}^{-\frac{t}{T_{1e}}}
\end{equation}
\begin{eqnarray}
&S_z(t) &= \mathrm{e}^{-\frac{t}{T_{2e}}}\left([S_z(0)+C_1J_z(0)+C_3J_y(0)]\cos{\omega_e t} + [S_y(0)+C_2J_z(0)+C_4J_y(0)]\sin{\omega_e t}\right)  \nonumber \\
&+& \mathrm{e}^{-\frac{t}{T_T}}\left([-C_1J_z(0)-C_3J_y(0)]\cos{\omega_h t} + [C_3J_z(0)-C_1J_y(0)]\sin{\omega_h t}\right)  
\end{eqnarray}
\begin{eqnarray}
&S_y(t) &= \mathrm{e}^{-\frac{t}{T_{2e}}}\left([S_y(0)+C_2J_z(0)+C_4J_y(0)]\cos{\omega_e t} - [S_z(0)+C_1J_z(0)+C_3J_y(0)]\sin{\omega_e t}\right) \nonumber \\
&+& \mathrm{e}^{-\frac{t}{T_T}}\left([-C_2J_z(0)-C_4J_y(0)]\cos{\omega_h t} + [C_4J_z(0)-C_2J_y(0)]\sin{\omega_h t}\right)  
\end{eqnarray}

Here
\begin{eqnarray}
C_1&=&\frac{\gamma}{2\tau_r}\left[\frac{1}{\gamma^2+(\omega_e+\omega_h)^2}+\frac{1}{\gamma^2+(\omega_e-\omega_h)^2} \right]=\frac{\gamma(\gamma^2 + \omega_e^2 + \omega_h^2)}{\tau_r(\gamma^2+(\omega_e+\omega_h)^2)(\gamma^2+(\omega_e-\omega_h)^2)}\nonumber \\
C_2&=&\frac{1}{2\tau_r}\left[\frac{\omega_e+\omega_h}{\gamma^2+(\omega_e+\omega_h)^2}+\frac{\omega_e-\omega_h}{\gamma^2+(\omega_e-\omega_h)^2} \right]=\frac{\omega_e(\gamma^2 + \omega_e^2 - \omega_h^2)}{\tau_r(\gamma^2+(\omega_e+\omega_h)^2)(\gamma^2+(\omega_e-\omega_h)^2)}\nonumber \\
C_3&=&\frac{1}{2\tau_r}\left[\frac{\omega_e+\omega_h}{\gamma^2+(\omega_e+\omega_h)^2}-\frac{\omega_e-\omega_h}{\gamma^2+(\omega_e-\omega_h)^2} \right]=\frac{\omega_h(\gamma^2 - \omega_e^2 + \omega_h^2)}{\tau_r(\gamma^2+(\omega_e+\omega_h)^2)(\gamma^2+(\omega_e-\omega_h)^2)}\nonumber \\
C_4&=&\frac{\gamma}{2\tau_r}\left[\frac{1}{\gamma^2+(\omega_e-\omega_h)^2}-\frac{1}{\gamma^2+(\omega_e+\omega_h)^2} \right]=\frac{2\gamma \omega_e \omega_h}{\tau_r(\gamma^2+(\omega_e+\omega_h)^2)(\gamma^2+(\omega_e-\omega_h)^2)}\nonumber \\
\end{eqnarray}

It is also useful for the subsequent calculations to introduce the following coefficients:
\begin{eqnarray}
D_1 &\equiv & \frac{C_1-C_4}{2}=\frac{\gamma}{2\tau_r(\gamma^2+(\omega_e+\omega_h)^2)}
\nonumber \\
D_2 &\equiv & \frac{C_1+C_4}{2}=\frac{\gamma}{2\tau_r(\gamma^2+(\omega_e-\omega_h)^2)}
\nonumber \\
D_3 &\equiv & \frac{C_2+C_3}{2}=\frac{(\omega_e+\omega_h)}{2\tau_r(\gamma^2+(\omega_e+\omega_h)^2)}
\nonumber \\
D_4 &\equiv & \frac{C_2-C_3}{2}=\frac{(\omega_e-\omega_h)}{2\tau_r(\gamma^2+(\omega_e-\omega_h)^2)}
\end{eqnarray}

 \section{Calculation of 3PE signals}
 
This section provides a procedure for calculating the 3-pulse spin dependent photon echo signal from a trion system. In this case, the specific form of the final and intermediate expressions for simplification regards some experimental conditions:
 \begin{itemize}
\item using linearly polarized pulses, polarized parallel to the direction of the magnetic field (H) or orthogonal to the magnetic field (V);
\item neglecting anisotropy of the $g$-factor of the electron and hole. This condition is equivalent to the specific orientation of the magnetic field relative to the crystal axes for which the effective magnetic field acting on electrons and holes is parallel or orthogonal to the external one~\cite{Trifonov-PRB2021}, as was done in the experiment. 
 \end{itemize}
 
The calculation procedure is divided into several serial steps corresponding to the experimental protocol. In this case, we track only those components of the density matrix, as well as their terms, which finally contribute to the detected signal. In the previous chapters, a detailed calculation of each of the possible steps was presented. The initial condition for each following step is the result of the previous one.
 
The calculation starts with the initial condition for the density matrix:
 \begin{equation}
 \rho(0) = \begin{pmatrix}
 1/2 & 0 & 0 &0 \\
 0 &1/2 & 0 &0 \\
 0 & 0 & 0 &0 \\
 0 & 0 & 0 &0 
 \end{pmatrix},
 \end{equation}
This corresponds to the ground state of the system with an equiprobable distribution of electron spin projections. Note, in a magnetic field this assumption is correct only if $\mu_B g_e B {\ll} k_B T $  which holds  for the experimental conditions used. Here, $k_B$ is the Boltzmann constant, and $T$ is the temperature.
 
 The calculation procedure has the following steps:
 \begin{enumerate}
\item The action of the first pulse (with H polarization). To calculate the final signal,  the knowledge of the off-diagonal components of the density matrix corresponding to the optical coherence is  sufficient.
\item Evolution of the off-diagonal density matrix elements assosiated with optical coherence  in the time interval $\tau_ {12}$.
\item The action of the second pulse (with H or V linear polarization). To calculate this signal, only the components of the density matrix which define the population and spin state of electrons and trions are sufficient. 
\item Evolution of elements of the density matrix which define the population and spin state of electrons and trions in the time interval $ \tau_{23} $.
\item Calculation of the detected signal in the required polarization configuration.
 \end{enumerate}

The intermediate results of calculations at each of the steps are presented below. Here, to identify the step, two-digit superscripts are introduced for the density matrix $\rho^{ij}$, where $j$ - corresponds to the ordinal number of the pulse that has acted (is acting) on the system, $i = (b, a)$ corresponds to the state before or after the action of the pulse.
 
 \paragraph{Step 1. Action of the first pulse.}
 
  \begin{equation}
 \rho^{b1} = \rho(0)
 \end{equation}  
\begin{eqnarray}
\rho_{13}^{a1}&=&\mathrm{i}\theta_{+1}^*L_{+1}K_{+1}\rho_{11}^{b1}, \nonumber \\
\rho_{24}^{a1}&=&\mathrm{i}\theta_{-1}^*L_{-1}K_{-1}\rho_{22}^{b1}, \nonumber \\
\rho_{31}^{a1}&=&(\rho_{13}^{a1})^*,  \nonumber \\ \rho_{42}^{a1}&=&(\rho_{24}^{a1})^*,\nonumber \\
\rho_{14}^{a1}&=&\rho_{23}^{a1}=\rho_{41}^{a1}=\rho_{32}^{a1} = 0.
\end{eqnarray} 

If the first pulse is linearly polarized along the direction of $B$ ($H$-polarization), then polarization angle of the first pulse is $\alpha_1 = 0$. We denote $K_{+1}=K_{-1}\equiv K_1$, $L_{+1}=L_{-1}\equiv L_1$, $\theta_{+1}=\theta_{01}\exp{-\mathrm{i} \alpha_1}=\theta_{01}$, ${\theta_{-1}}=\theta_{01}\exp{\mathrm{i} \alpha_1}=\theta_{01}$. 

In this way $$\rho_{13}^{a1}=\rho_{24}^{a1}=\frac{\mathrm{i}}{2}\theta_{01}^*L_{1}K_{1}.$$

\paragraph{Step 2. Evolution of off-diagonal elements between 1st and 2nd pulses.}

\bea
\rho_{13}^{b2} &=& \frac{\mathrm{i}}{2}\theta_{01}^*L_{1}K_{1} (\cos(\omega_e \tau_{12}/2)\cos(\omega_h \tau_{12}/2)+\sin(\omega_e \tau_{12}/2)\sin(\omega_h \tau_{12}/2))\mathrm e^{-\tau_{12}/T_2}  \nonumber\\
\rho_{14}^{b2} &=& \frac{1}{2}\theta_{01}^*L_{1}K_{1}(\cos(\omega_e  \tau_{12}/2)\sin(\omega_h  \tau_{12}/2)-\sin(\omega_e  \tau_{12}/2)\cos(\omega_h  \tau_{12}/2))\mathrm e^{- \tau_{12}/T_2} \nonumber\\
\rho_{23}^{b2} &=&  \frac{1}{2}\theta_{01}^*L_{1}K_{1}(\sin(\omega_e \tau_{12}/2)\cos(\omega_h \tau_{12}/2)-\cos(\omega_e \tau_{12}/2)\sin(\omega_h \tau_{12}/2))\mathrm e^{-\tau_{12}/T_2}  \nonumber\\
\rho_{24}^{b2} &=& \frac{\mathrm{i}}{2}\theta_{01}^*L_{1}K_{1}(\sin(\omega_e \tau_{12}/2)\sin(\omega_h \tau_{12}/2)+\cos(\omega_e \tau_{12}/2)\cos(\omega_h \tau_{12}/2))\mathrm e^{-\tau_{12}/T_2} 
\eea 

Using simple trigonometric algebra one can simplify equations above as follows:

\bea
\rho_{13}^{b2} &=& \frac{\mathrm{i}}{2}\theta_{01}^*L_{1}K_{1} \cos([\omega_e -\omega_h] \tau_{12}/2)\mathrm e^{-\tau_{12}/T_2} \nonumber\\
\rho_{14}^{b2} &=&   \frac{1}{2}\theta_{01}^*L_{1}K_{1}\sin([\omega_e-\omega_h] \tau_{12}/2)\mathrm e^{-\tau_{12}/T_2} \nonumber\\
\rho_{23}^{b2} &=&   \frac{1}{2}\theta_{01}^*L_{1}K_{1}\sin([\omega_e-\omega_h] \tau_{12}/2)\mathrm e^{-\tau_{12}/T_2}  \nonumber\\
\rho_{24}^{b2} &=& \frac{\mathrm{i}}{2}\theta_{01}^*L_{1}K_{1}\cos([\omega_e -\omega_h] \tau_{12}/2)\mathrm e^{-\tau_{12}/T_2}  
\eea 

 \paragraph{Step 3. Action of second pulse.}
 
 The second pulse is polarized linearly along direction $\alpha_2$. This means that $K_{+2}=K_{-2}\equiv K_2$, $L_{+2}=L_{-2}\equiv L_2$, $\theta_{+2}=\theta_{02}\mathrm e^{-\mathrm{i} \alpha_2}$, ${\theta_{-2}}=\theta_{02}\mathrm e^{\mathrm{i} \alpha_2}$. 
 
 For further calculations of the stimulated echo signal, we only need terms proportional to $\theta_{\pm 1}\theta_{\pm 2}^*$ (or to $\theta_{\pm 1}^*\theta_{\pm 2}$), because only they contribute to the {non-linear coherent response determined by the $E_1^*E_2E_3$ contribution which corresponds to the rephasing scheme and consequently the photon echo signal~\cite{Cho-JPC1992}. For convenience and clarity, we take only the terms proportional to $\theta_{\pm1}\theta_{\pm2}^*$. In the final expressions, one can take everything into account by adding complex conjugate expressions. It is also worth noting that the complete expressions for diagonal elements are real. After the second pulse action: 
 
\bea
\rho_{11}^{a2}&\sim &-\mathrm{i}L_2\theta_{02}\mathrm e^{\mathrm{i} \alpha_2}K_2^*\rho_{31}^{b2}
\nonumber\\
\rho_{33}^{a2}&\sim &\mathrm{i}L_2\theta_{02}\mathrm e^{\mathrm{i} \alpha_2}K_2^*\rho_{31}^{b2}
\nonumber\\
\rho_{22}^{a2}&\sim &-\mathrm{i}L_2\theta_{02}\mathrm e^{-\mathrm{i} \alpha_2}K_2^*\rho_{42}^{b2}
\nonumber\\
\rho_{44}^{a2}&\sim &\mathrm{i}L_2\theta_{02}\mathrm e^{-\mathrm{i} \alpha_2}K_2^*\rho_{42}^{b2}
\nonumber\\
\rho_{12}^{a2}&\sim &-\mathrm{i}\theta_{02}\mathrm e^{\mathrm{i} \alpha_2}L_2K_2^*\rho_{32}^{b2}
\nonumber\\
\rho_{21}^{a2}&\sim &-\mathrm{i}\theta_{02}\mathrm e^{-\mathrm{i} \alpha_2}L_2K_2^*\rho_{41}^{b2}
\nonumber\\
\rho_{34}^{a2}&\sim &\mathrm{i}\theta_{02}\mathrm e^{-\mathrm{i} \alpha_2}L_2K_2^*\rho_{32}^{b2}
\nonumber\\
\rho_{43}^{a2}&\sim &\mathrm{i}\theta_{02}\mathrm e^{\mathrm{i} \alpha_2}L_2K_2^*\rho_{14}^{b2}
\eea 
 
Substituting $\rho_{31}^{b2}, \rho_{42}^{b2}, \rho_{32}^{b2}, \rho_{41}^{b2}$, we get:

\bea
\rho_{11}^{a2} &\sim & - A_{2p}\mathrm{e}^{\mathrm{i} \alpha_2} \cos([\omega_e -\omega_h] \tau_{12}/2) 
\nonumber\\
\rho_{33}^{a2} &\sim & A_{2p}\mathrm{e}^{\mathrm{i} \alpha_2} \cos([\omega_e -\omega_h] \tau_{12}/2) 
\nonumber\\
\rho_{22}^{a2} &\sim & - A_{2p}\mathrm{e}^{-\mathrm{i} \alpha_2}  \cos([\omega_e -\omega_h] \tau_{12}/2) 
\nonumber\\
\rho_{44}^{a2} &\sim & A_{2p}\mathrm{e}^{-\mathrm{i} \alpha_2}  \cos([\omega_e -\omega_h] \tau_{12}/2) \nonumber\\
\rho_{12}^{b2} &\sim & -\mathrm{i}A_{2p}\mathrm{e}^{\mathrm{i} \alpha_2}  \sin([\omega_e-\omega_h] \tau_{12}/2)  \nonumber\\
\rho_{21}^{b2} &\sim &  -\mathrm{i}A_{2p}\mathrm{e}^{-\mathrm{i} \alpha_2} \sin([\omega_e-\omega_h] \tau_{12}/2) \nonumber\\
\rho_{34}^{b2} &\sim & \mathrm{i}A_{2p}\mathrm{e}^{-\mathrm{i} \alpha_2} \sin([\omega_e-\omega_h] \tau_{12}/2)  \nonumber\\
\rho_{43}^{b2} &\sim & \mathrm{i}A_{2p}\mathrm{e}^{\mathrm{i} \alpha_2}  \sin([\omega_e-\omega_h] \tau_{12}/2)  \nonumber\\
\eea 

Here $A_{2p}=\frac{\theta_{01}\theta_{02}}{2}L_1L_2K_1^*K_2^*\mathrm e^{-\tau_{12}/T_2}$.

Note, here for simplicity we also  have omitted multipliers $\cos{(\omega_0\tau_{12})}$ and $\sin{(\omega_0\tau_{12})}$. These multipliers describe formation of spectral grating responsible for the stimulated photon echo signal formation (see details in Suppl. Mat. of Ref.~\cite{Langer-NatPhot2014}).

\bea
S_z^{a2} &\sim & -\mathrm{i}A_{2p} \sin{(\alpha_2)}  \cos([\omega_e-\omega_h] \tau_{12}/2)  
\nonumber\\
J_z^{a2} &\sim & \mathrm{i}A_{2p} \sin{(\alpha_2)}  \cos([\omega_e-\omega_h] \tau_{12}/2)  
\nonumber\\
S_y^{a2} &\sim & \mathrm{i}A_{2p} \sin{(\alpha_2)}  \sin([\omega_e-\omega_h] \tau_{12}/2)  
\nonumber\\
J_y^{a2} &\sim & \mathrm{i}A_{2p} \sin{(\alpha_2)}  \sin([\omega_e-\omega_h] \tau_{12}/2)  
\nonumber\\
S_x^{a2} &\sim &-\mathrm{i}A_{2p} \cos{(\alpha_2)}   \sin([\omega_e-\omega_h] \tau_{12}/2)  
\nonumber\\
J_x^{a2} &\sim &\mathrm{i}A_{2p}\cos{(\alpha_2)}   \sin([\omega_e-\omega_h] \tau_{12}/2) 
\nonumber\\
n_e^{a2} &\sim & -2A_{2p} \cos{(\alpha_2)}   \cos([\omega_e-\omega_h] \tau_{12}/2)  
\nonumber\\
n_h^{a2} &\sim & 2A_{2p} \cos{(\alpha_2)}   \cos([\omega_e-\omega_h] \tau_{12}/2)  
\nonumber\\
\eea

Thus, the spin state after the action of two pulses is:
 
HH ($\alpha_1=\alpha_2=0$)
 \begin{eqnarray}
 S_z^{a2}=S_y^{a2}=0 = J_z^{a2}= J_y^{a2} \\
 S_x^{a2} = - J_x^{a2}  \sim \sin{(\frac{\omega_e-\omega_h}{2}\tau_{12})}  \mathrm{e}^{-\frac{\tau_{12}}{T_2}} 
 \label{eq:SG-X}
  \end{eqnarray}
 
 HV($\alpha_1=0, \alpha_2=\pi/2$)
 \begin{eqnarray}
 S_z^{a2} = - J_z^{a2} \sim  \cos{(\frac{\omega_e-\omega_h}{2}\tau_{12})} \mathrm{e}^{-\frac{\tau_{12}}{T_2}}  \\
 \label{eq:SG-Z}
 S_y^{a2} = J_y^{a2} \sim -\sin{(\frac{\omega_e-\omega_h}{2}\tau_{12})} \mathrm{e}^{-\frac{\tau_{12}}{T_2}}  \\
 \label{eq:SG-Y}
 S_x^{a2} = J_x^{a2} = 0
 \end{eqnarray}

\paragraph{Step 4. Evolution of  density matrix elements in population and spin representation.}
According to section III and omitting common multiplier $-\mathrm{i}A_{2p}$
\begin{equation}
n_T^{b3} \sim 2\mathrm{i}\cos{(\alpha_2)} \cos([\omega_e-\omega_h] \tau_{12}/2) \mathrm{e}^{-\frac{\tau_{23}}{\tau_r}}
\end{equation}
\begin{equation}
J_x^{b3} \sim -\mathrm{e}^{-\frac{\tau_{23}}{T_T}}\cos{(\alpha_2)}\sin([\omega_e-\omega_h] \tau_{12}/2)
\end{equation}
\begin{equation}
J_y^{b3} \sim -\mathrm{e}^{-\frac{\tau_{23}}{T_T}}\sin{(\alpha_2)}\sin([\omega_e-\omega_h] \tau_{12}/2-\omega_h \tau_{23})
\end{equation}
\begin{equation}
J_z^{b3} \sim -\mathrm{e}^{-\frac{\tau_{23}}{T_T}}\sin{(\alpha_2)}\cos([\omega_e-\omega_h] \tau_{12}/2-\omega_h \tau_{23})
\end{equation}

\begin{equation}
n_e^{b3} \sim -2\mathrm{i}\cos{(\alpha_2)} \cos([\omega_e-\omega_h] \tau_{12}/2) \mathrm{e}^{-\frac{\tau_{23}}{\tau_r}}
\end{equation}
\begin{equation}
S_x^{b3} \sim  \mathrm{e}^{-\frac{\tau_{23}}{T_{1e}}}\cos{(\alpha_2)}\sin([\omega_e-\omega_h] \tau_{12}/2)
\end{equation}
\begin{eqnarray}
&S_z^{b3} &\sim \mathrm{e}^{-\frac{\tau_{23}}{T_{2e}}}\sin{(\alpha_2)}\left(\cos([\omega_e-\omega_h] \tau_{12}/2+\omega_e \tau_{23})\right.
\nonumber \\
&-&D_2\cos([\omega_e-\omega_h] \tau_{12}/2-\omega_e \tau_{23}) 
- D_1\cos([\omega_e-\omega_h] \tau_{12}/2+\omega_e \tau_{23})
 \nonumber \\
&-&\left.D_3\sin([\omega_e-\omega_h] \tau_{12}/2+\omega_e \tau_{23})+D_4\sin([\omega_e-\omega_h] \tau_{12}/2-\omega_e \tau_{23})\right)  
\nonumber \\
&+& \mathrm{e}^{-\frac{\tau_{23}}{T_T}}\sin{(\alpha_2)}\left(C_1\cos([\omega_e-\omega_h] \tau_{12}/2-\omega_h \tau_{23})+C_3\sin([\omega_e-\omega_h] \tau_{12}/2-\omega_h \tau_{23})\right)  
\end{eqnarray}
\begin{eqnarray}
&S_y^{b3} &\sim \mathrm{e}^{-\frac{\tau_{23}}{T_{2e}}}\sin{(\alpha_2)}\left(-\sin([\omega_e-\omega_h] \tau_{12}/2+\omega_e \tau_{23})\right.
\nonumber \\
&-&D_3\cos([\omega_e-\omega_h] \tau_{12}/2+\omega_e \tau_{23}) 
- D_4\cos([\omega_e-\omega_h] \tau_{12}/2-\omega_e \tau_{23})
 \nonumber \\
&-&\left.D_2\sin([\omega_e-\omega_h] \tau_{12}/2-\omega_e \tau_{23})+D_1\sin([\omega_e-\omega_h] \tau_{12}/2+\omega_e \tau_{23})\right)  
\nonumber \\
&+& \mathrm{e}^{-\frac{\tau_{23}}{T_T}}\sin{(\alpha_2)}\left(C_2\cos([\omega_e-\omega_h] \tau_{12}/2-\omega_h \tau_{23})+C_4\sin([\omega_e-\omega_h] \tau_{12}/2-\omega_h \tau_{23})\right)  
\end{eqnarray}

 \paragraph{Step 5. Calculation of detected signal in required polarization}
 
 The spin state of the system at the moment of arrival of the third pulse determines the detected 3-pulse spin-dependent long-lived photon echo signal. Below we provide the explanation and the final expression for the detected signal.
 
 The third pulse is also linearly polarized, therefore $\theta_{+3}=\theta_{03}\mathrm{e}^{-\mathrm{i}\alpha_3}$, $\theta_{-3}=\theta_{03}\mathrm{e}^{\mathrm{i}\alpha_3}$, $K_{+3}=K_{-3}\equiv K_3$, $L_{+3}=L_{-3}\equiv L_3$.
For calculation of stimulated echo signal, we only need terms proportional to $\theta_{\pm3}^*$, because before we have chosen terms proportional to $\theta_{\pm1}\theta_{\pm2}^*$. This combination of elements ($\theta_{\pm1}\theta_{\pm2}^*\theta_{\pm3}^*$) will give the radiation corresponding to the photon echo.

In accordance with Eqs. (10) - (12): 
 \bea 
 \rho_{13}^{a3} &\sim & -\mathrm{i} \theta_{+3}^*K_{+3}L_{+3} (\rho_{33}^{b3}-\rho_{11}^{b3})=-\mathrm{i} \theta_{03}K_{3}L_{3} (\rho_{33}^{b3}-\rho_{11}^{b3}) \mathrm{e}^{\mathrm{i} \alpha_3}, \nonumber \\
 \rho_{24}^{a3} &\sim & -\mathrm{i}\theta_{-3}^*K_{-3}L_{-3} (\rho_{44}^{b3}-\rho_{22}^{b3})= -\mathrm{i}\theta_{03}K_{3}L_{3} (\rho_{44}^{b3}-\rho_{22}^{b3}) \mathrm{e}^{-\mathrm{i} \alpha_3}, \nonumber \\
 \rho_{14}^{a3} &\sim & \mathrm{i}(\theta_{-3}^*K_{+3}L_{-3}\rho_{12}^{b3}-\theta_{-3}^*K_{-3}L_{+3}\rho_{34}^{b3})=\mathrm{i}\theta_{03}K_{3}L_{3} (\rho_{12}^{b3}\mathrm{e}^{-\mathrm{i} \alpha_3}-\rho_{34}^{b3}\mathrm{e}^{\mathrm{i} \alpha_3}), \nonumber \\
 \rho_{23}^{a3} &\sim & \mathrm{i}(-\theta_{-3}^*K_{+3}L_{-3}\rho_{43}^{b3}+\theta_{-3}^*K_{-3}L_{+3}\rho_{21}^{b3})=\mathrm{i}\theta_{03}K_{3}L_{3} (-\rho_{43}^{b3}\mathrm{e}^{-\mathrm{i} \alpha_3}+\rho_{21}^{b3}\mathrm{e}^{\mathrm{i} \alpha_3}), 
 \eea
 
The detected signal is proportional to the macroscopic polarization of the system at the time moment $\tau_ {12}$ after the action of the third pulse:
 \begin{equation}
 P_{3PE} \sim \mathrm{e}^{-\mathrm{i} \alpha_d} \rho_{13}^{d} + \mathrm{e}^{\mathrm{i} \alpha_d} \rho_{24}^{d} + c.c.,
 \end{equation}
 where $\alpha_d$ is the angle of detection linear polarization, $\rho^{d}$ is the density matrix in moment of PE formation. 
 
 In experiments we used H detection polarization $\alpha_d =0$ (co-polarized with first pulse):
 \begin{equation}
 P_{3PE} \sim \rho_{13}^{d} + \rho_{24}^{d} \sim \left[(\rho_{13}^{a3} + \rho_{24}^{a3})\cos(\frac{[\omega_e - \omega_h] \tau_{12}}{2})-\mathrm{i}(\rho_{14}^{a3} + \rho_{23}^{a3})\sin(\frac{[\omega_e - \omega_h] \tau_{12}}{2})\right]\mathrm{e}^{-\tau_{12}/T_2}+c.c.,
 \end{equation} 

Omitting common multiplier ($ -\mathrm{i} \theta_{03}^*K_{3}L_{3}\mathrm{e}^{-\tau_{12}/T_2} $):
 
  \bea 
 P_{3PE} &\sim&  ( \rho_{33}^{b3}  \mathrm{e}^{\mathrm{i} \alpha_3}  -\rho_{11}^{b3} \mathrm{e}^{\mathrm{i} \alpha_3} + \rho_{44}^{b3}\mathrm{e}^{-\mathrm{i} \alpha_3} -\rho_{22}^{b3} \mathrm{e}^{-\mathrm{i} \alpha_3} )\cos([\omega_e - \omega_h] \tau_{12}/2) \nonumber \\
 &+& \mathrm{i}(\rho_{21}^{b3}\mathrm{e}^{\mathrm{i} \alpha_3} + \rho_{12}^{b3}\mathrm{e}^{-\mathrm{i} \alpha_3}-\rho_{43}^{b3}\mathrm{e}^{-\mathrm{i} \alpha_3} -\rho_{34}^{b3}\mathrm{e}^{\mathrm{i} \alpha_3}) \sin([\omega_e - \omega_h] \tau_{12}/2) 
 \eea

Finally, we have:
 \bea 
 P_{3PE} &\sim& \theta_{03}K_{3}L_{3}\mathrm{e}^{-\tau_{12}/T_2} [ -\mathrm{i}(n_T^{b3}-n_e^{b3})\cos(\alpha_3)+2(J_z^{b3}-S_z^{b3})\sin(\alpha_3)]\cos([\omega_e - \omega_h] \tau_{12}/2) \nonumber \\
 &+& [ 2(S_x^{b3}-J_x^{b3})\cos(\alpha_3)-2(J_y^{b3}+S_y^{b3})\sin(\alpha_3)] \sin([\omega_e - \omega_h] \tau_{12}/2) +c.c.
 \eea

\section{3PE signal in HHHH and HVVH polarization configurations}

The 3PE signal has two components which have short and long decay time with increasing $\tau_{23}$. The first one is decaying with trion lifetime $\tau_r$ and its spin lifetime $T_T$. In case of long trion spin relaxation time $T_h\gg\tau_r$ this signal decays with trion lifetime only. The long-lived component decays with spin relaxation time of resident electrons. The 3PE signal in HHHH polarization configuration is:
   
    \begin{eqnarray}
  P_{HHHH} &\sim&  -A_{3p}\mathrm{e}^{-\frac{2 \tau_{12}}{T_2}}\left(2\mathrm{e}^{-\frac{\tau_{23}}{\tau_r}} \cos^2{\left(\frac{\omega_e-\omega_h}{2} \tau_{12}\right) } + \mathrm{e}^{-\frac{\tau_{23}}{T_T}}\sin^2{\left(\frac{\omega_e-\omega_h}{2} \tau_{12} \right)}\right. 
  \nonumber \\
  &+& \left. \mathrm{e}^{-\frac{\tau_{23}}{T_{1e}}}\sin^2{\left(\frac{\omega_e-\omega_h}{2} \tau_{12}\right) }\right)
 \end{eqnarray}

Here $A_{3p}=\mathrm{i}\theta_{01}\theta_{02}\theta_{03}L_1L_2L_3K_1^*K_2^*K_3$
and the LSPE {signal} is the last term on the right hand side. For example, for resonant ($\Delta=0$) $\pi/2$-pulses $A_{3p}=-\mathrm{i}/8$. 

Stimulated PE signal in HVVH polarization configuration is:   
 \begin{eqnarray}
P_{HVVH} &\sim & A_{3p}\mathrm{e}^{-\frac{\tau_{23}}{T_{2e}}}(\cos{[(\omega_e-\omega_h) \tau_{12}+\omega_e \tau_{23}]}-D_1\cos{[(\omega_e-\omega_h) \tau_{12}+\omega_e \tau_{23}]}
\nonumber \\
&-&D_3\sin{[(\omega_e-\omega_h) \tau_{12}+\omega_e \tau_{23}]}-D_2\cos{(\omega_e \tau_{23})}-D_4\sin{(\omega_e \tau_{23})})\nonumber \\
&+& A_{3p}\mathrm{e}^{-\frac{\tau_{23}}{T_T}}(\cos{[(\omega_e-\omega_h) \tau_{12}-\omega_h \tau_{23}]}+D_1\cos{[(\omega_e-\omega_h) \tau_{12}-\omega_h \tau_{23}]}
\nonumber \\
&+&D_3\sin{[(\omega_e-\omega_h) \tau_{12}-\omega_h \tau_{23}]}+D_2\cos{(\omega_h \tau_{23})}+D_4\sin{(\omega_h \tau_{23})})+c.c.
  \end{eqnarray}
One can rewrite the last expression as follows: 
\begin{eqnarray}
&& P_{HVVH} \sim  A_{3p}[\mathrm{e}^{-(\frac{2\tau_{12}}{T_2}+\frac{\tau_{23}}{T_{2e}})}(p_e\cos{(\omega_e \tau_{23}+(\omega_e-\omega_h) \tau_{12})}+q_e\sin{(\omega_e \tau_{23}+(\omega_e-\omega_h)\tau_{12})})
\nonumber \\
&+& \mathrm{e}^{-(\frac{2\tau_{12}}{T_2}+\frac{\tau_{23}}{T_T})}(p_h\cos{(\omega_h \tau_{23}-(\omega_e-\omega_h)\tau_{12})}+q_h\sin{(\omega_h \tau_{23}-(\omega_e-\omega_h)\tau_{12})})]+c.c.\equiv
\nonumber \\
&\equiv &A_{3p}\mathrm{e}^{-\frac{2 \tau_{12}}{T_2}}[\mathrm{e}^{-\frac{\tau_{23}}{{T_{2e}}}}r_e\cos{(\omega_e \tau_{23}+(\omega_e-\omega_h)\tau_{12}-\phi_e)}+\mathrm{e}^{-\frac{\tau_{23}}{T_T}}r_h\cos{(\omega_h \tau_{23}-(\omega_e-\omega_h)\tau_{12}-\phi_h)}]+
\nonumber \\
&+&c.c.
\label{Phvvh_final}
  \end{eqnarray}
Here
\begin{eqnarray}
p_e&= & 1-D_1-D_2\cos[(\omega_e-\omega_h)\tau_{12}]+D_4\sin[(\omega_e-\omega_h) \tau_{12}]
\nonumber \\
q_e&= & -D_3-D_2\sin[(\omega_e-\omega_h)\tau_{12}]-D_4\cos[(\omega_e-\omega_h) \tau_{12}]
\nonumber \\
p_h&= & 1+D_1+D_2\cos[(\omega_e-\omega_h)\tau_{12}]+D_4\sin[(\omega_e-\omega_h) \tau_{12}]
\nonumber \\
q_h&= & -D_3-D_2\sin[(\omega_e-\omega_h)\tau_{12}]+D_4\cos[(\omega_e-\omega_h) \tau_{12}]
  \end{eqnarray}

  \begin{equation}
  r_e=\sqrt{p_e^2+q_e^2}, r_h=\sqrt{p_h^2+q_h^2}
 \end{equation}
  \begin{equation}
  \cos\phi_e=p_e/r_e, \sin\phi_e=q_e/r_e, \cos\phi_h=p_h/r_h, \sin\phi_h=q_h/r_h.
 \end{equation}
 
 In a strong enough magnetic field ($\gamma \ll \omega_e,\omega_h,\omega_e\pm\omega_h$), all $D_i \to 0$, $(p_e, p_h) \to 1$, $(q_e, q_h) \to 0$ and SPE signal is:
\begin{eqnarray}
P_{HVVH} &\sim & A_{3p}\mathrm{e}^{-\frac{2 \tau_{12}}{T_2}}[\mathrm{e}^{-\frac{\tau_{23}}{T_{2e}}}\cos{(\omega_e \tau_{23}+(\omega_e-\omega_h) \tau_{12})}
\nonumber \\ 
&+&\mathrm{e}^{-\frac{2 \tau_{12}}{T_2}}\mathrm{e}^{-\frac{\tau_{23}}{T_T}}\cos{(\omega_h \tau_{23}-(\omega_e-\omega_h) \tau_{12})}].
  \end{eqnarray}

 \section{List of used physical quantities}

$g_{e(h)}$ is electron (hole) in-plane $g$ factor.

$\omega_{e(h)} = g_{e(h)} \mu_B B/ \hbar$ is electron (hole) Larmor precession angular frequency.

$\mu_B$ is Bohr's magneton.

$B$ is {external} magnetic field strength.

$\omega$ is central {optical} frequency of laser pulses.

$\omega_0$ is frequency of optical transition.

$\Delta = \omega-\omega_0$ is detuning between laser and resonance frequencies. 

$T_2$ is decay time of optical coherence.

$T_1$ is energy relaxation {time} from {optically} excited state.

$T_{1e(1h)}$ is longitudinal electron (hole) spin relaxation time.

$T_{2e(2h)}$ is transverse electron (hole) spin relaxation time.

$\tau_r$ is trion lifetime.

$T_T = T_h \tau_r/(T_h + \tau_r)$ is trion spin lifetime, where isotropic hole relaxation is assumed with $T_h = T_{2h} = T_{1h}$.

$d$ is dipole moment of optical transitions

$\theta$ is pulse area.

{$\Omega$ is generalized Rabi frequency.

\end{document}